# Specific aspects of internal corrosion of nuclear clad made of Zircaloy

J.B. Minne[1a], L. Desgranges[2b], V. Optasanu[3c],
N. Largenton[1d], L. Raceanu[3e] and T. Montesin[3f]

[1]EDF R&D, Les Renardières, Route de Sens, 77818 Moret sur Loing cedex, France

[2]CEA/DEN/DEC/SESC/LLCC, CEA Cadarache, 13115 St Paul lez Durance, France

[3]ICB, Université de Bourgogne, UMR 5209 CNRS, 9 Avenue A. Savary, Dijon, France

[a]jean-baptiste.minne@edf.fr, [b]lionel.desgranges@cea.fr, [c]virgil.optasanu@u-bourgogne.fr,
[d]nathalie.largenton@edf.fr, [e]laura.raceanu@u-bourgogne.fr, [f]tony.montesin@u-bourgogne.fr

**Abstract:** In PWR, the Zircaloy based clad is the first safety barrier of the fuel rod, it must prevent the dispersion of the radioactive elements, which are formed by fission inside the $UO_2$ pellets filling the clad. We focus here on internal corrosion that occurs when the clad is in tight contact with the $UO_2$ pellet. In this situation, with temperature of 400 °C on the internal surface of the clad, a layer of oxidised Zircaloy is formed with a thickness ranging from 5 to 15 µm. In this paper, we will underline the specific behaviour of this internal corrosion layer compared to wet corrosion of Zircaloy. Simulations will underline the differences of stress field and their influences on corresponding dissolved oxygen profiles. The reasons for these differences will be discussed as function of the mechanical state at inner surface of the clad which is highly compressed. Differences between mechanical conditions generated by an inner or outer corrosion of the clad are studied and their influences on the diffusion phenomena are highlighted.

**Introduction**

Understanding the mechanisms of corrosion of zirconium alloys in the nuclear industry is an important issue because it is one of the parameters that affect the service life of fuel rods. It is generally accepted that, as long as the growing oxide layer remains protective, the diffusion of oxygen vacancies is the rate-limiting mechanism for zirconium oxidation; the oxide growth occurs at the metal-oxide interface and leads to a parabolic kinetic law. Several types of corrosion have been studied in order to represent the different oxidizing environments of the clad and lead to other mechanisms. Uniform waterside corrosion in reactor operating conditions (from 280 to 330 °C) is one of the most widely reviewed [1,2].

Internal corrosion at inner clad surface is another specific zirconium alloy oxidation in operating conditions which has been less studied. However, the evolution of this layer is of importance since the formation of an oxide layer leads to improve transfer of both heat and mechanical contact forces across the pellet/clad gap. As the burn-up increases, the gap between the $UO_2$ pellet and the clad decreases progressively to the closure. A tight pellet/clad contact then occurs, which generally corresponds to the initiation of the internal corrosion layer growth [3,4]. Moreover, this contact is related to a quick increase of the hoop stress in the clad, and also to a pellet/clad contact pressure.

Since the importance of stress generation in the oxide and the metal was often highlighted for a better understanding of the mechanisms involved in waterside corrosion [5], this paper underlines the influences of mechanical conditions generated by the inner and outer clad corrosion on the diffusion phenomena. We present first specific aspects of the growth kinetics and of the stress state of internal corrosion layers compared to waterside corrosion. Then, the use of a model which couples stress and oxygen diffusion enables us to appreciate the feedback of mechanical effects on the dissolved oxygen profiles.

**Waterside corrosion vs. internal corrosion of Zircaloy clad**

In water, for PWR conditions, clad oxidation occurs in two stages, referred to as pre- and post-transition regimes. In pre-transition regime, a black oxide film develops. It is characterized by a dense and protective layer of mixed tetragonal and monoclinic zirconia, the latter is the stable phase in this temperature range. Kinetics is rather closer to a cubic or quartic law than a parabolic one. The thickness of the layer controls the oxidation rate and is characterized by high compression stresses (1-2 GPa) near the oxide/metal interface.

After 2-3 µm, the layer degrades, cracks and becomes porous and a tetragonal-to-monoclinic transformation of zirconia occurs. The kinetic law switches to a pseudo-linear post-transition regime, which is constituted by the cyclic succession of several cubic regimes and accelerates the corrosion rate. The impact of the stress level in the zirconia appears here to be of great importance. The transformation from metal to oxide produces a high volume change, as the so-called Pilling-Bedworth ratio is 1.56 and leads to high residual stress in both the metal and the oxide layer. These stresses are responsible for these cracks in the waterside oxide layer and are often correlated to phase transformation in the zirconia [7].

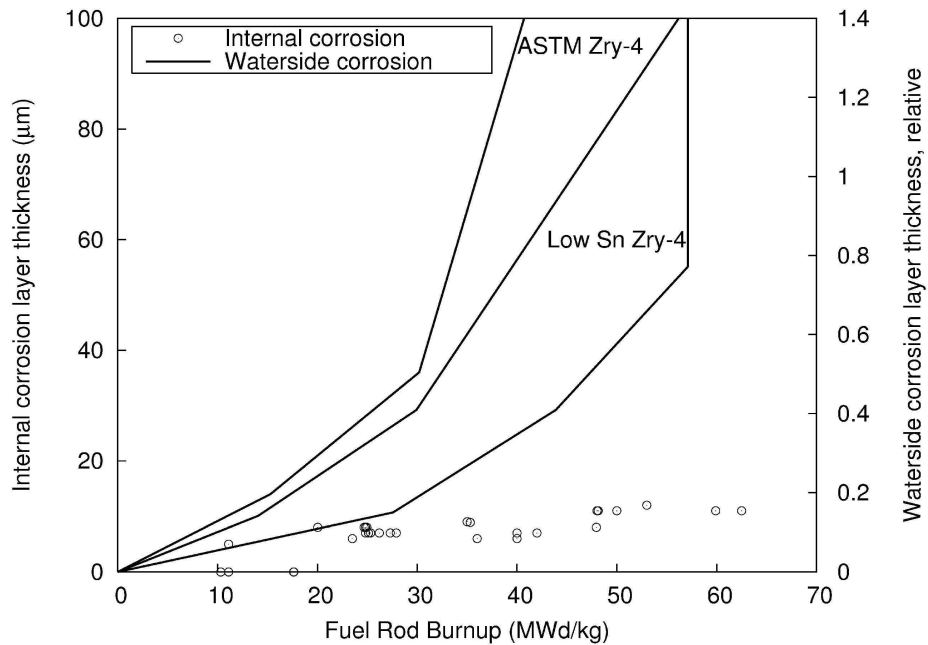

Figure 1: Comparison of internal [2] and waterside corrosion thickness [3] in PWR for $UO_2$/MOX fuel and Zry-4/low Sn Zry-4 clad as function of the fuel rod burn-up.

In the case of corrosion of the inner clad surface, conditions of oxide formation and growth kinetics are totally different, in comparison with waterside corrosion (see Fig. 1). Thus, internal zirconia comes as a fully tetragonal and dense layer, though this phase is stable at temperatures much higher than those of internal clad surface (340-400 °C) and at higher pressure. However, tetragonal zirconia could be induced at these temperatures by high stresses (> 4 GPa), or is stabilized by the addition of others elements (Y, Mg), or is produced by fission-induced phase transformation [7]. In this last case, the pressure of the high concentrations of interstitials (fission products) produced by elastic collisions would cause a denser material, thus a more symmetrical structure of zirconia.

It is generally reported that internal oxidation is observed for closed radial gap between the pellet and the clad, in operating conditions [3,4]. As the oxide layer grows first discontinuously in small patches, we can assume that the establishing of a local tight pellet/clad contact is needed to enable the oxygen diffusion from pellet to clad, which coincides with the mechanical conditioning of the rod. The growth rate is composed of 3 steps. The first 5 µm oxide thicknesses grow very quickly and correspond to the penetration length of fission products from pellet to clad. Moreover, it is well established that diffusion under radiation is drastically enhanced due to the ballistic migration of atoms [8].

The layer grows then slowly up to 10-12 µm. Diffusion could be governed by a diffusion rate-limiting mechanism, like the "pre-transition" regime of waterside corrosion. An estimation of the parabolic diffusion rate constant leads to a value of the order of magnitude of $3.10^{-19}$ m²/s, which is coherent with common values for this range of temperature [9]. At high burn-up, this thickness does not increase significantly, it does not exceed 15 µm and non-connected porosity is only observed near the oxide/metal interface after 3 cycles of irradiation. No residual gap is observed, which is related to a chemical interaction and to a good mechanical hang-up between the oxide layer and the pellet.

The development of a wrinkled surface can be observed at the interface between the pellet and the corrosion layer, linked with the high burn-up structure at the edge of the pellet. These wrinkles can be strongly marked with grooves and folds; the layer thickness can periodically double, which favour the good cohesion between the clad and the pellet.

Since the stresses were proved to have a non negligible feedback effect on the diffusion rate of oxygen in the metal [10], it would be interesting to know how the chemical diffusion process is affected by the stress field generated by the bonding and by the pressure of the pellet. For this purpose, we introduce the model used for our FEM simulations and the adopted methodology.

**Model description of zirconium oxidation**

The background theory which gives the influence of mechanical stress on the chemical diffusion was widely exposed in [10-12], and is based on previous works [13]. We recall here the main point used in our simulations. In the volume, the matter flux relative to the diffusing species (oxygen in our case), is given by a modified Fick's law [14] in which the diffusion coefficient $D$ depends on the mechanical stress:

$$D = D_0 \left[ 1 - \frac{M_0 \eta_{ij} \sigma_{ij}}{RT} \right] \quad (1)$$

where $M_0$ is the molar mass of the zirconium, $\eta_{ij}$ are the chemical expansion tensor coefficients, $c$ is the oxygen concentration which is dissolved in zirconium, $\sigma$ is the mechanical stress, $T$ is the temperature and $R$ is the universal gas constant. $\eta_{ij}$ are defined as follows:

$$\eta_{ij} = \rho \frac{\partial \varepsilon_{ij}^{ch.ox}}{\partial c} \quad (2)$$

where $\rho$ is the density of the zirconium and $\varepsilon_{ij}^{ch.ox}$ are the chemical strain tensor coefficients, which are assumed to be isotropic in the metal [17]. For zirconia, chemical expansion coefficients are evaluated according to the chemical strain coefficients determined elsewhere [15]. Table 1 shows the parameters used for the model. The solids behaviours are elastic and isotropic.

| $T = 400\ °C$ | $D$ (m²/s) [9] | $\eta_{ij}$ (m³/kg) | $E$ (GPa) | $\nu$ |
|---|---|---|---|---|
| UO$_2$ | | | 175 | 0.31 |
| ZrO$_2$ | | $\begin{cases} \eta_r = 116.4 \times 10^{-5} \\ \eta_\theta = \eta_z = 1.08 \times 10^{-5} \end{cases}$ | 100 | 0.25 |
| Zry-4 | $1.966 \times 10^{-19}$ | $3.55 \times 10^{-5}$ | 95 | 0.31 |

Table 1: Parameters used for the model.

We use here a strong coupling between concentration and mechanical stress. The stress locally modifies the diffusion coefficient and the oxygen concentration induces modifications in chemical strains, which produces additional mechanical stress. We assume that only oxygen is able to diffuse and produces one single phase change, that the chemical reaction of this phase change is always at the thermodynamic equilibrium, that the system is isothermal and that there is a perfect adhesion between the oxide and the metal, i.e., that there is a mechanical continuity at the oxide/metal interface.

**Numerical simulation of oxygen dissolution into the metal**

The numerical simulations were performed with the finite element CAST3M code applied to a 2 dimension axi-symmetric mesh which contains 4 domains (see Fig. 2): the edge of the pellet (500 µm length), the internal zirconia layer (5 and 15 µm thickness), the clad (570 µm length) and the external zirconia layer (2 µm thickness, which is the effective mean thickness that is involved in the pre-transition regime of external corrosion, and 6 µm thickness). The solubility limit of oxygen in the clad is set to $c^{met/ox} = 464$ kg/m$^3$ [16]. The UO$_2$ pellet is under internal pressure $p_{int}$ and the clad is under external pressure $p_{ext}$ with $p_{int} = p_{ext} = 155$ bars.

We simulated the diffusion of oxygen in the clad for several thicknesses of internal and external zirconia layers during $3\times10^7$ s, which corresponds to a 10 µm penetration length of oxygen in the metal. No mobility of the interfaces was hence simulated.

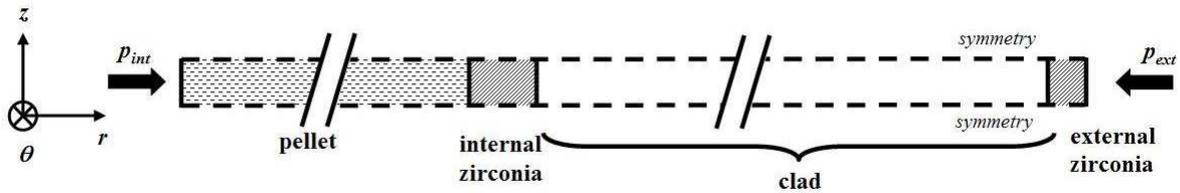

Figure 2: Geometry of the numerical model

We focus here on the oxygen diffusion in the α-Zr(O) phase and on the bulk pressure in the clad, defined as $(\sigma_r+\sigma_\theta+\sigma_z)/3$. Comparisons between stress-diffusion coupling and standard Fick's diffusion are made on inner and outer sides of the clad for several zirconia thicknesses.

**Results**

Results are given in figures 3 and 4. The effect of the stress-diffusion coupling model is remarkable compared to the Fick's diffusion: the oxygen concentration can locally double (see Fig. 3). No significant differences between the inner and outer sides of the clad are noticed for both oxygen concentration profiles and bulk pressure.

As the adopted mechanical model is elastic, calculated compressive stresses in the α-Zr(O) phase are very high due to the isotropic chemical expansion coefficient. However, they are in the same order of magnitude of measured hoop stresses for Zircaloy-4 plate oxidized at 400°C in steam [18]. The bulk pressure in the clad, outside the α-Zr(O) phase, increases with the inner and outer oxide thickness, since the Young modulus of zirconia is greater than the clad's one, which tends to stiffen the clad.

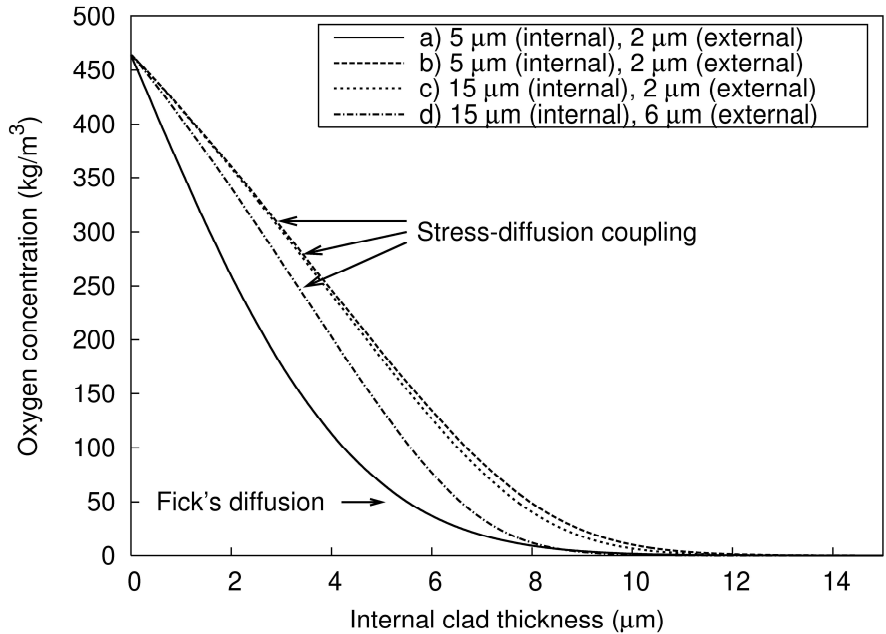

Figure 3: Oxygen concentration at the inner side of the clad.
a) Fick's diffusion. b), c) and d) Stress-diffusion coupling for different values of thickness of internal and external zirconia.

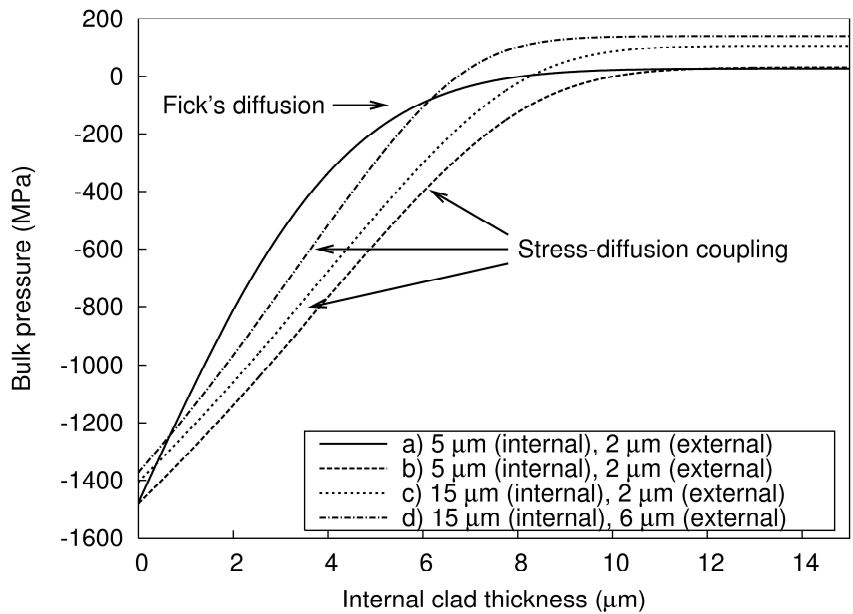

Figure 4: Bulk pressure (right) at the inner side of the clad.
a) Fick's diffusion. b), c) and d) Stress-diffusion coupling.

**Discussion**

Our results show that, the more internal and external thickness increases, the more the bulk pressure increases in the clad and the less the oxygen diffuses in the metal. According to figure 4, the bulk pressure in the clad, outside the α-Zr(O) phase, increases with the inner and outer oxide thickness, since the Young modulus of zirconia is greater than the clad's one, which tends to stiffen the clad. As the oxide layer increases, on inner and outer clad surface, compressive stresses decreases, which is related to a decrease of the oxygen diffusion coefficient in the $\alpha$-Zr(O) phase and explains the difference of concentration profiles. Actually, according to equation 1, increasing the tensile stress or decreasing the compressive stress allows the diffusion coefficient in the metal to decrease. Then, the feedback of the stress state on the corrosion kinetics lowers the rate diffusion of oxygen in the metal.

In service conditions, waterside corrosion thickness can reach a few dozen micrometers at high burn-up. However, only the last 2-3 µm oxide layer thickness in front of the zirconia/metal interface is involved in the rate-limiting mechanism for the clad oxidation. The rest of the layer is destabilized and degraded. This leads to the loss of protective properties of the zirconia layer and to stress relaxation in the clad. We can then conclude that the external penetration profile of oxygen does not change a lot during the lifetime of the rod even if oxidation kinetics is accelerated.

Unlike the outer zirconia, internal corrosion layer is formed under irradiation and fission products collisions. The stabilization of the tetragonal phase is then induced by radiation damage, and then no phase transformation can occur. For that case, the internal penetration profile of oxygen could evolve, as the oxide layer remains dense and protective.

## VII Conclusion

The present work is a first approach of the modelling of the internal corrosion of nuclear clad. The phenomenology of the growths of internal and waterside corrosion, in service conditions, and the specific stress state of the pellet/clad system have been analysed in order to simulate the impact of mechanical stress on oxygen diffusion in the clad with the help of a stress-diffusion model. The main point of our results is that dissolution of oxygen in the clad can be interpreted in terms of external zirconia destabilization and internal zirconia stabilization. The future works will simulate the growth of zirconia layers in presence of mechanical stress.